# Breaking Ion Clusters: Size Asymmetry for Faster Ion Transport in Polymer Electrolytes

Ganesh K Rajahmundry and  Tarak K Patra[*]

Department of Chemical Engineering, Indian Institute of Technology Madras, Chennai, TN- 600036, India

Center for Atomistic Modeling and Materials Design, Indian Institute of Technology Madras, Chennai, TN-600036, India

## Abstract

Solid polymer electrolytes (SPEs) are ion-containing solid materials composed of a polymer matrix that enables ionic transport while maintaining the mechanical stability. The conventional wisdom is that for a high ion concentration, ions microphase separate from the polymer matrix, resulting in poor conductivity. Instead, we show that a high ion size ratio promotes better mixing of the ions with the polymer matrix. Under these conditions, the ion-dipole moment interaction dominates over the ion-ion interaction and improves the ion dispersion in the polymer matrix. The ion size ratio is thus a key to tailor the properties of these materials with immediate relevance to the development of SPEs for energy storage devices.

**Graphical Abstract**

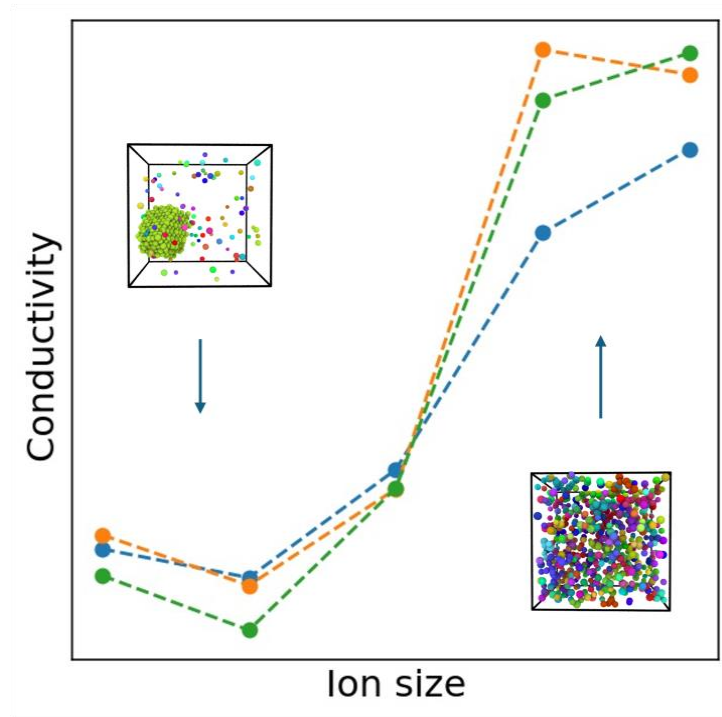

Solid polymer electrolytes (SPEs) are mixtures of salts and polymers such as LiTFSI filled Poly(ethylene oxide) (PEO) and LiClO$_4$ filled polymethyl methacrylate (PMMA). They are promising materials for energy storage applications. They can provide better electrochemical and mechanical stability to batteries over the currently used liquid electrolytes.[1–7] The added salts in SPEs dissociate into ions and they migrate through the polymer matrix, thereby imparting ion conductivity to the material. However, a key limitation of SPEs is their low room-temperature ionic conductivity (~10$^{-4}$ S/cm),[5,8] which is much lower than that of liquid electrolytes. In addition, enhancing the energy density of SPEs remains challenging, as increasing the salt concentration often leads to ion aggregation.[9–12] This mocrophase separation disrupts the uniform ion transport pathways and lower the overall conductivity of SPEs. Therefore, ongoing research efforts are focused on developing new design concepts and material engineering strategies to enhance the ion distribution and subsequent ionic conductivity of SPEs.[1,13–15]

Studies over the last decade suggest that the ion transport in SPEs is intricately linked to the molecular architecture of the polymer and the resulting microstructure.[16–19] However, the effects of ion topology on the overall conductivity of an SPE remain under appreciated. Interestingly, a recent experiment indicates that the ionic conductivity of an SPE exhibits a monotonic relationship to the cation size.[20] However, the microscopic origin of this ion-size-dependent conductivity remains poorly understood. We hypothesize that the relative ion size ratio acts as a key control parameter for the phase behaviour of the electrolyte, determining whether the system favors dispersion of ions or their aggregation. This microstructural transition, in turn, governs the overall ionic conductivity. Here, we test this hypothesis and bridge the gap between ion size effects and resulting ion transport in SPEs.

We use coarse-grained molecular dynamics simulations (CGMD) to study the above hypotheses, as they have been successfully used to establish correlations between molecular scale structural properties and conductivity in ion-doped polymers as well as ion-containing polymers.[21–31] These simulations have significantly advanced our understanding of the phase behavior and ionic conductivity of SPEs. Towards this end, in our recent CGMD study, we investigate a limiting case of SPEs where the polymer matrix was assumed to possess no dipole moment.[32] While this simplified model provided important insights, it does not fully capture the complexity of experimentally relevant SPEs, many of which exhibit significant dipole moments arising from polar functional groups. These dipoles are expected to interact strongly with free ions in the system, thereby altering the balance between dipole–ion and ion–ion interactions. Building upon our earlier framework, we extend the model to explicitly incorporate dipole moments within the polymer matrix. This allows us to systematically examine how dipole–ion coupling competes with ion–ion correlations, and how this interplay governs the emergent phase behaviour of SPEs. By doing so, we aim to develop a more comprehensive molecular-level understanding of the factors that control ion organization and transport in dipolar polymer electrolytes.

We systematically vary the ion size ratio and ion concentration and compute the ion distribution and ion conductivity. These data are utilized to construct a phase diagram of the model SPE as a function of the ion size ratio and ion concentration. We find that at low ion concentrations, the ions are well-dispersed and uniformly mixed within the polymer matrix, leading to a homogeneous phase. However, as the ion concentration increases, the system undergoes a crossover from an ion-dispersed to an ion-aggregated state, the nature of which strongly depends on the ion size ratio. Lower ion size ratio promotes phase segregation and the formation of ion-rich domains, whereas the higher ion size ratio favors ion dispersion in the polymer matrix. As a consequence, at high ion concentrations, the ionic conductivity increases significantly with the ion size ratio. This enhancement arises because larger ion size asymmetry suppresses ion aggregation and facilitates the formation of more continuous ion-transport pathways within the polymer matrix. Consequently, systems with greater ion size disparity exhibit more efficient ion mobility and improved conductivity. Based on this study, we expect that the combination of an increment in ion size ratio and ion concentration can supress ion clustering, leading to better ion transport in SPEs. Overall, we demonstrate the complex correlations between ion size ratio and phase behaviour of an SPE that dictate its conductivity.

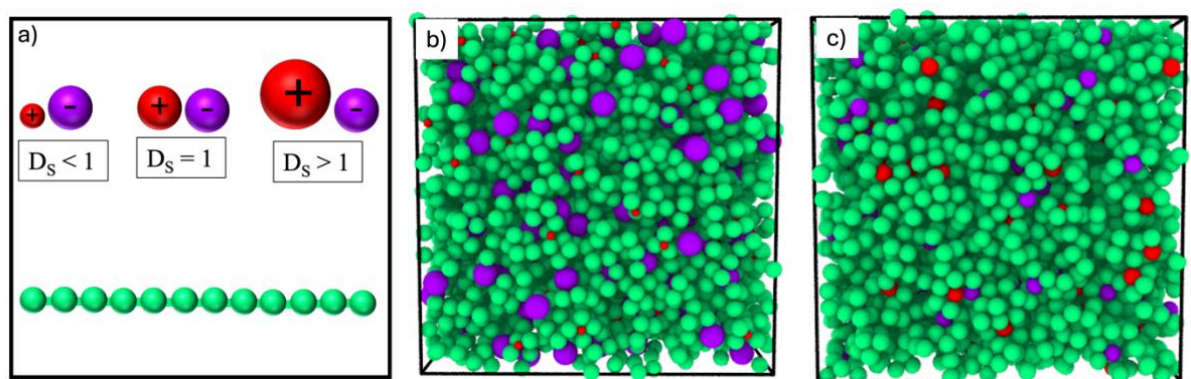

*Figure 1: SPE Model: A schematic representation of the model system with three types of ion size ratios considered is shown in (a). The polymer is made of 40 monomers. Two equilibrium MD snapshots correspond to $D_S$ =0.3 and $D_S$=1 are shown in (b) and (c), respectively. The ion concentration in (b) and (c) is $\rho_{ion} = 2\%$.*

A schematic representations of the model system is shown in Figure 1a, wherein ions and monomers of a polymer are considered as CG beads. We employ a polymer model whose effective dipole moment closely matches that of the PEO, and the charges assigned to the CG ions are chosen to closely approximate the effective charges of Li$^+$ and TFSI$^-$, as parameterized and reported by Wheatle et al.[28] This ensuring a realistic representation of electrostatic and dipolar interactions in the CG polymer electrolyte model. In this model, two adjacent monomers of a polymer is connected by the finitely extensible non-linear elastic (FENE) potential of the form $E_{FENE}(r_{ij}) = -0.5kr_0^2 \ln\left[1 - \left(\frac{r_{ij}}{r_0}\right)^2\right]$. Here, $r_0 = 1.5\sigma$ and k = 30$\varepsilon$ with $\sigma$ and $\epsilon$ being the size of a monomer, and the cohesive interaction strength between a pair of monomers, respectively. In addition, a pair of monomers interacts via the Weeks-Chandler-Andersen(WCA) potential of the form $V(r) = 4\epsilon\left[\left(\frac{\sigma}{r}\right)^{12} - \left(\frac{\sigma}{r}\right)^6\right]$, which is truncated and shifted to zero at $r = 1.12\sigma$. Ions are also model as spherical CG particles that interact via the WCA potential. The anion size is $d_- = 1\sigma$. The cation size $(d_+)$ is varied from $0.5\sigma$ to $1.5\sigma$. The excluded volume interaction between ion and monomer is also modeled using the WCA potential. The Lorentz Berthelot mixing rule $\sigma_{ij} = (\sigma_i + \sigma_j)/2$ is used for the effective size of the WCA interactions between particles with dissimilar sizes. The cohesive interaction strength is $\epsilon$ for all pairs of excluded volume interactions. Each monomer poses a dipole moment, and the dipole-dipole interaction is modeled as $E_{pp}(r) = \left[\frac{1}{r^3}(\vec{p}_i \cdot \vec{p}_j) - \frac{3}{r^3}(\vec{p}_i \cdot \vec{r})(\vec{p}_j \cdot \vec{r})\right]$. The dipole-ion interaction is represented by $E_{qp}(r) = \frac{q}{r^3}(\vec{p} \cdot \vec{r})$. Here, $\vec{p}$ is the dipole moment of a monomer, and $q$ is the charge of an ion. The electrostatic interaction between a pair of ion is represented via the Coulomb potential of the form $E_{coul}(r) = \frac{q_i q_j}{4\pi\varepsilon_0 \varepsilon_r r}$. Here, the $\varepsilon_0$ is the permittivity of vacuum, $\varepsilon_r$ is the relative dielectric constant of the medium. The $q_i$ and $q_j$ are charges of i$^{th}$ and j$^{th}$ ions, respectively. The CG ion charge is $|q| = 9.83$, and the CG monomer dipole moment is $p = 1.7$. The ion-ion, ion-dipolar and dipole-dipole interactions are truncated at a cut-off distance of $5\sigma$, with the Ewald summation[33] is used to account for contributions beyond this limit. We consider 120 polymer chains, each consist of 40 monomers. We vary the ion concentration $(\rho_{ion})$, which is defined as $\rho_{ion} = \frac{Number\ of\ ion\ pairs}{Number\ of\ monmers} \times 100$, from 2% to 8.33%. The ion size ratio (D$_s$), which is defined as $D_s = \frac{d_+}{d_-}$, varies from 0.5 to 1.5. All simulations are performed using an isothermal-isobaric (*NPT*) ensemble with a reduced pressure $\left(\frac{P\sigma^3}{\epsilon}\right) = 0.72$ and a reduced temperature $\left(\frac{k_B T}{\epsilon}\right) = 0.97$ for $10^7$ steps of equilibration followed by a production run of $4 \times 10^7$ steps with an integration time step of $0.005\tau$. Here, $T$, $P$ and $k_B$ are the temperature, pressure and Boltzmann constant, respectively. The temperature and pressure are mentioned by the Nosé-Hover thermostat and barostat, respectively. All the simulations are conducted using the LAMMPS MD package.[34] Two representative equilibrium configurations of the system are shown in Figure 1b and 1c.

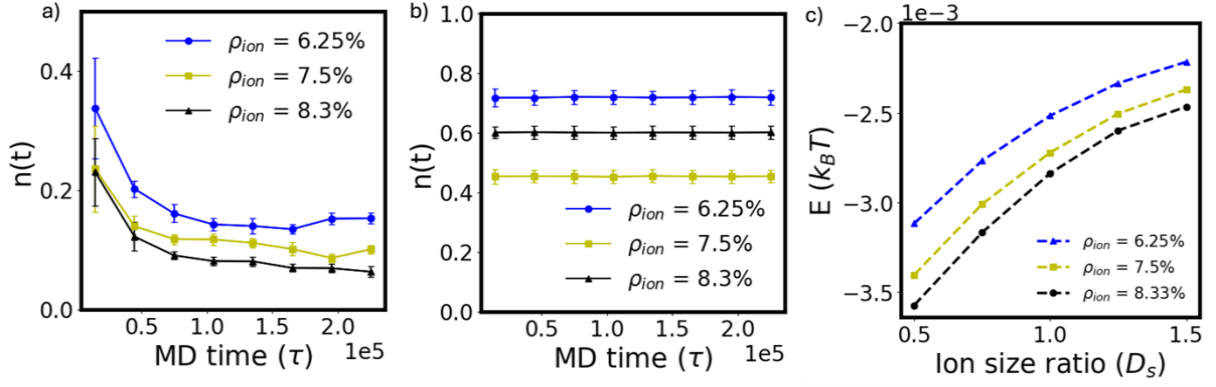

*Figure 2: Free ion fraction n(t) is shown as a function of MD time in (a) and (b) for Ds= 0.75 and Ds=1.25, respectively. The (a) and (b) correspond to phase separation an dispersion, respectively. The pair energy of the system is plotted as a function of $D_s$ for three ion concentrations in (c).*

We quantify the fractions of ions that form aggregates and those that remain free in the polymer matrix during the equilibration. These ion aggregates are quantified using the standard cluster analysis procedure from simulation data.[28,32,35–37] If two or more ions are considered to be part of the cluster if their center-center distance is less than mean of their size. The free ion factor *n(t)* is defined as the ratio of free ions to the total ions in the system. The n(t) is plotted as a function of MD time in Figure 2a and b. It decreases gradually and reaches a plateau as a function of time for $D_s$=0.75 (Figure 2a), which indicate clustering of ions in the polymer matrix. Whereas for $D_s$=1.25, the *n(t)* does not change appreciably over the MD run, (Figure 2b), which corresponds to ion dispersion in the polymer matrix. Therefore, the free ion distribution distinguishes the two phases of the system. Furthermore, to gain deeper insight into the underlying thermodynamic origin of the observed phase change as a function of ion size ratio, we compute the pairwise interaction energy (*E*) during the production run, which is shown in Figure 2c. The results reveal a systematic decrease in the magnitude of *E* of the equilibrated system with increasing ion size ratio, indicating a weakening of ion–ion correlations. This reduction in cohesive interaction energy facilitates enhanced ion–polymer affinity, thereby promoting a more homogeneous ion dispersion within the polymer matrix. In other words, larger ion size ratio mitigates aggregation tendencies and drives the system toward a mixed phase. We construct a phase diagram based on these MD simulation results in Figure

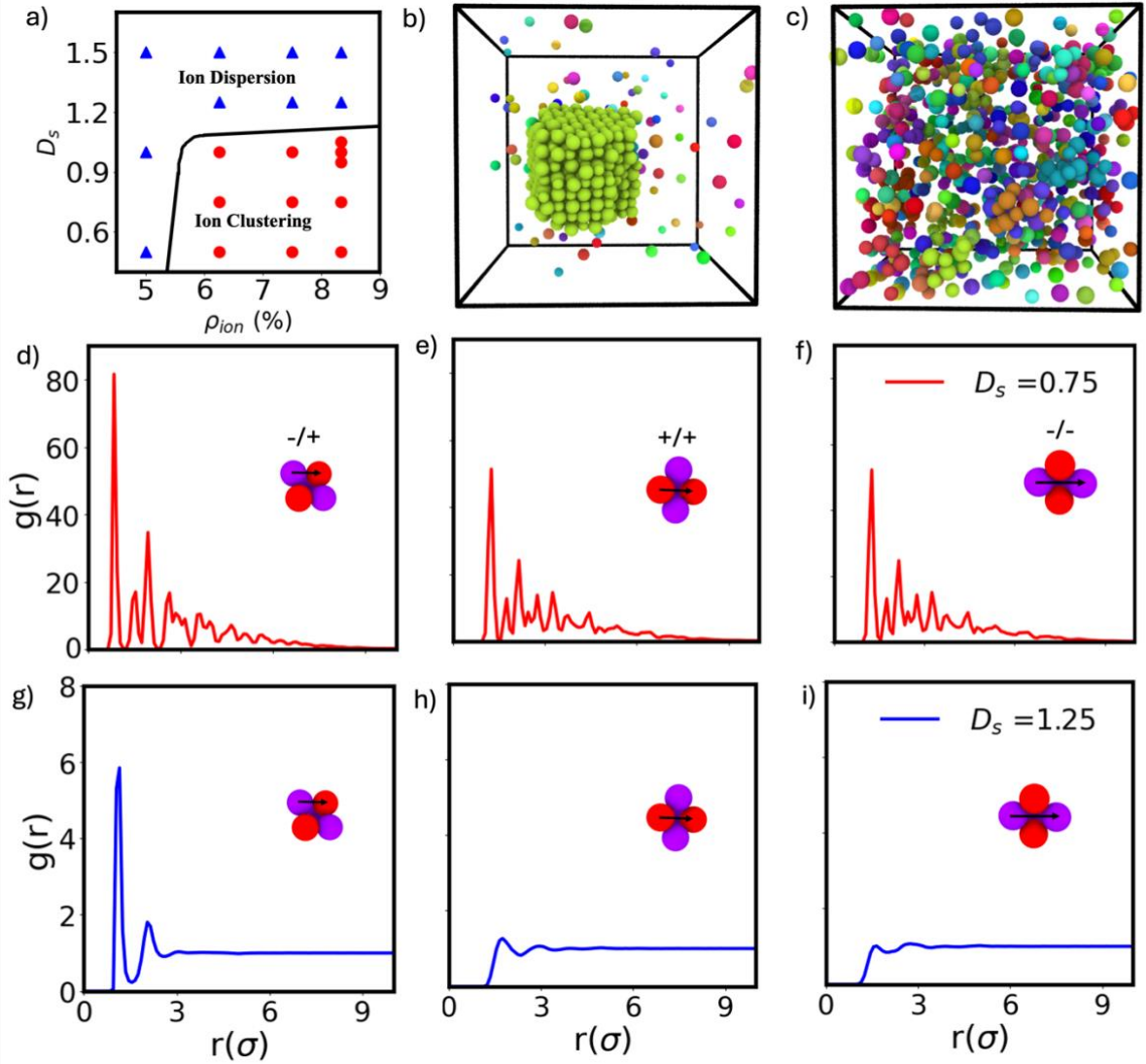

*Figure 3: The phase diagram is shown in (a) wherein the solid line is guide to eyes. Blue triangles represent ion dispersion and red circles represent ion clustering. MD snapshots of the system are shown in (b) and (c) for two representative cases of ion clustering and ion dispersion, respectively. Polymers are deleted from the snapshots (b ,c) for better visual inspection of the ion distribution. Ions are color-coded based on their cluster identity, with all ions within a given cluster shown in the same color. The RDF for cation-anion , cation-cation, anion-anion pairs are shown in panels (d, g), (e, h), and (f, i), respectively. The ion size ratios are $D_s=0.75$ and $D_s=1.25$ for panels (d-f) and (g-i), respectively.*

3a. At low ion concentrations, the ions remain well dispersed regardless of the ion size ratio. However, at higher ion concentrations ( $\rho_{ion} > \sim 5\%$), two distinct regimes emerge—an ion-dispersed phase and an ion-cluster phase depending on the $D_s$. Two representative MD snapshots of the system for *$D_s = 0.75$* and *$D_s = 1.25$* are shown in Figures 3(b) and 3(c), respectively, for $\rho_{ion} =$ 7.5%. These MD snapshots illustrate the ion distribution corresponding to the two distinct phases. The fraction of ions belonging to the largest cluster is varied from 25 to 90% of the total ions in the ion-aggregation region of the Figure 3a. whereas, the fraction of ions belonging to the largest cluster is less that 4% of the total ions in the ion-dispersed region of the Figure 3a. In Figure 3b and c, ions are color-coded according to their cluster identities, such that distinct colors are assigned to ions belonging to different clusters. The ion-ion pair correlation functions are shown in Figure 3(d-f) and Figure 3(g-i) for $D_s = 0.75$ and $D_s = 1.25$, respectively. They suggest that ion-ion pair correlations are stronger when they form large clusters. However, the pair correlations weaken as they disperse for higher ion size ratio.

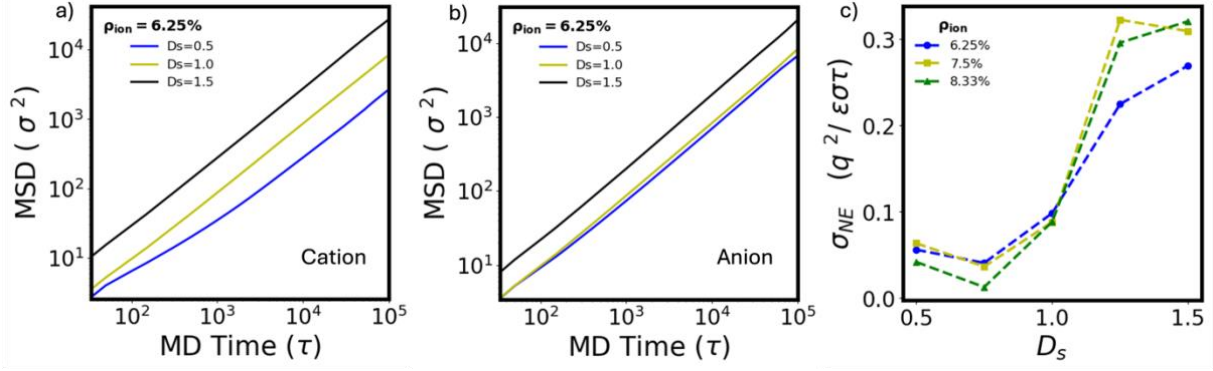

*Figure 2: Mean square displacement of an ion is plotted as a function of time in (a) and (b) for cation and anion, respectively. The Nernst-Einstein conductivity is shown as a function of the ion size ratio ($D_s$) in (c) for three different ion concentrations.*

We calculate the ion conductivity using the Nernst-Einstein relation which can be written as $\sigma_{NE} = \frac{e^2}{Vk_BT}(N_+z_+^2 D_+ + N_-z_-^2 D_-)$,[38,35,28] where $e$ is the elementary charge, $k_B$ is the Boltzmann's constant, and $V$ is the volume of the system. Here, $D, z$ and $N$ are diffusivity, charge and number of ions, with the subscripts + and - corresponding to the cation and anion, respectively. The diffusivity is calculated from the slope of a mean squared displacement (MSD) function of an ion in its diffusive regime, which is written as $D = \frac{1}{6t}\sum_{n=1}^{N} \lim_{t\to\infty} \langle |r_i(t) - r_i(0)|^2 \rangle$. The MSD curves for few representative cases are shown in Figure 4a and b for a high ion concentration. The ion conductivity is plotted as a function of ion pair size ratio in Figure 4c for different ion concentrations. The results indicate that with increasing ion size ratio, the ionic conductivity exhibits an abrupt enhancement coinciding with the transition from an ion-clustered to an ion-dispersed regime. This sharp increase suggests a strong coupling between the microstructural organization of ions and macroscopic charge transport. In particular, the disruption of ion clusters and the emergence of a more homogeneous ion distribution facilitate the formation of continuous conductive pathways through the polymer matrix. Therefore, the ionic conductivity of the SPE is highly sensitive to its phase behavior, emphasizing the critical role of nanoscale morphology in determining ion transport efficiency.

In summary, we employ CGMD simulations to elucidate how the ion size ratio governs their spatial distribution in polymers with strong dipole moments. Our results reveal that the ionic conductivity in such polymer media exhibits a systematic dependence on the underlying phase behavior. Specifically, at elevated ion concentrations, smaller ion size ratios promote their microphase separation, whereas larger size ratios favor homogeneous ion dispersion. These findings identify the ion size ratio as a critical design parameter in SPEs, enabling rational control over their morphology and transport properties through molecular-level tuning.

## Acknowledgements

This work is made possible by financial support from the SERB through a core research grant (CRG/2022/006926). This research uses the resources of the Center for Nanoscience Materials, Argonne National Laboratory, which is a DOE Office of Science User Facility supported under the Contract DE-AC02-06CH11357. G.K.R acknowledges Prime Minister Research Fellowship (PMRF) supported by the Ministry of Education, Government of India.